\documentclass[review]{elsarticle}
\usepackage{hyperref}
\usepackage{graphics} 
\usepackage[labelfont=bf,justification=raggedright,singlelinecheck=false]{caption}
\journal{Journal of \LaTeX\ Templates}

\begin{document}

\begin{frontmatter}

\title{A Technique for Estimating the Absolute Gain of a Photomultiplier Tube}

%\author{.........\fnref{myfootnote}}
%\address

\author[icrr]{M. Takahashi\corref{mycorrespondingauthor}}
\author[konan]{Y. Inome}
\author[konan]{S. Yoshii}
\author[Tokyo]{A. Bamba}
\author[yamagata]{S. Gunji}
\author[icrr]{D. Hadasch}
\author[icrr]{M. Hayashida}
\author[ibaraki]{H. Katagiri}
\author[kyoto]{Y. Konno}
\author[kyoto]{H. Kubo}
\author[tokai]{J. Kushida}
\author[icrr]{D. Nakajima}
\author[yamagata]{T. Nakamori}
\author[saitama]{T. Nagayoshi}
\author[tokai]{K. Nishijima}
\author[kyoto]{S. Nozaki}
\author[icrr]{D. Mazin}
\author[kyoto]{S. Mashuda}
\author[mip]{R. Mirzoyan}
\author[icrr]{H. Ohoka}
%\author[nagoya]{A. Okumura}
\author[tokushima]{R. Orito}
\author[icrr]{T. Saito}
\author[icrr]{S. Sakurai}
\author[yamagata]{J. Takeda}
%\author[ibaraki]{D.V. Tan}
\author[icrr]{M. Teshima}
\author[saitama]{Y. Terada}
\author[yamagata]{F. Tokanai}
%\author[konan]{T. Yamamoto}
\author[konan]{T. Yamamoto\corref{mycorrespondingauthor}}
\cortext[mycorrespondingauthor]{Corresponding author}
\ead{tokonatu@konan-u.ac.jp}
\author[ibaraki]{T. Yoshida}

\address[icrr]{Institute of Cosmic Ray Research, University of Tokyo, Kashiwa, Chiba 277-8582, Japan}
\address[konan]{Faculty of Science and Engineering, Konan University, Kobe 658-8501, Japan}
\address[Tokyo]{Department of Physics, University of Tokyo,  Bunkyo-ku, Tokyo 113-0033, Japan} 
\address[mip]{Max Planck Institute for Physics, D-80805 Munich Germany}
%\address[nagoya]{Institute for Space-Earth Environmental Research, Nagoya University, Nagoya, Aichi 464-8602, Japan}
%\address[nagoya]{ISEE, Nagoya University, Nagoya, Aichi 464-8602, Japan}
\address[tokushima]{ Institute of Socio- Arts and Science, Tokushima University, Tokushima 770-8502, Japan }
%\address[ibaraki]{ College of Science, Ibaraki University, 2-1-1, Bunkyo, Mito 310-8512, Japan }
\address[ibaraki]{ College of Science, Ibaraki University, Mito 310-8512, Japan }
\address[kyoto]{Department of Physics, Kyoto University, Kyoto 606-8502, Japan}
\address[tokai]{Department of Physics, Tokai University, Hiratsuka, Kanagawa 259-1292, Japan }
\address[yamagata]{Department of Physics, Yamagata University, Yamagata 990-8560, Japan}
\address[saitama]{Department of Physics, Saitama University, Saitama 338-8570, Japan}

\begin{abstract}
Detection of low-intensity light relies on the conversion of photons to photoelectrons, which are then multiplied and detected as an electrical signal. To measure the actual intensity of the light, one must know the factor by which the photoelectrons have been multiplied. To obtain this amplification factor, we have developed a procedure for estimating precisely the signal caused by a single photoelectron. 
The method utilizes the fact that the photoelectrons conform to a Poisson distribution. 
The average signal produced by a single photoelectron can 
then 
be estimated from the number of noise events, without requiring analysis of the distribution of the signal produced by a single photoelectron. The signal produced by one or more photoelectrons can be estimated experimentally without any assumptions. This technique, and an example of the analysis of a signal from a photomultiplier tube, are described in this study.
\end{abstract}

\begin{keyword}
%\texttt{elsarticle.cls}\sep \LaTeX\sep Elsevier \sep template
%\MSC[2010] 00-01\sep  99-00
Photomultiplier, PMT, photoelectron, photon detector %, Light sensor 
\end{keyword}

\end{frontmatter}

%%%%%\linenumbers

\section{Introduction}
%\subsection{Signal from a single photoelectron}

Light is quantized as photons. If it is very weak, the photons can be counted, which is equivalent to measuring the intensity of the weak light. Since photons are neutral particles, it is difficult to detect them directly. In general, a photon detector converts photons to electrons, which are thus called ''photoelectrons.'' The so-called ''quantum efficiency'' of a detector is proportional to the efficiency of this conversion from photons to electrons, whereas the overall efficiency of detecting a photon is called the ''photo-detection efficiency'' $[1]$. %\cite{Hamamatsu}. 

Measurements with high photo-detection efficiency and high precision are important for astronomical observations, such as 
the Large Size Telescopes (LSTs) in the Cherenkov Telescope Array (CTA) $[2]$. %\cite{CTA}. 
The goal of the CTA project is to construct the largest observatory of gamma-ray-imaging, atmospheric Cherenkov telescopes, devoted to observations of high-energy photons, with energies ranging from 20 GeV to 300 TeV.
The sensitivity in the lowest energy range will be dominated by four LSTs located at the center of the array. 
To be able to detect low-energy gamma rays,
each LST has a large mirror (23 m in diameter) and a high-sensitivity camera. The focal-plane instrument in the camera has to measure weak light with high precision. This requirement brought about the study reported in the present paper.
%%%%%%%%%%%%%%%%%%%%%%%%%%%%%%%%%%

%\subsection{Method for detecting a single photoelectron}

The electrical charge of a photoelectron is $1.6 \times 10^{-19} $ C. Since this value is extremely small, photoelectrons must be multiplied to be detected as an electrical signal.
%%%
Photomultiplier tubes (PMTs) are widely used to measure the intensity of such weak light. Photoelectrons are amplified in the PMT and extracted as an electrical signal. Since the amplification is a stochastic process, the electrical signal fluctuates. The fluctuations make the distribution of the signal produced by a single photoelectron wider, depending on the quality of the PMT. In this study, we term the photoelectron signal distribution produced by a single photoelectron ''1 PESD.'' Distributions of two or more photoelectrons result in the superposition of 1 PESD signals. The number of photoelectrons in a signal can thus be determined by dividing the signal by the average value of 1 PESD, and the number of photons can then be determined from the number of photoelectrons divided by the photon-detection efficiency. The goal of this paper is to estimate the average value of a 1 PESD signal.

With the improvement of PMT quality, accurate calibration and quality control become important for experiments such as LST in the CTA. We have developed a new technique to measure precisely 1 PESD. In this technique, we first estimate the number of zero-photoelectron events in the data. 
Then we estimate 1 PESD through an iteration analysis.
%The number of events that contain more than zero photoelectrons can then be calculated by utilizing the fact that the distribution of photoelectrons in an event is a Poisson distribution. 
%
%
Based on this analysis, the average intensity of the signal produced by a single photoelectron, together with its statistical error, can be estimated from the number of zero-photoelectron events, and the amplification factor for a single photoelectron can be estimated. In other words, the amplification factor and its error can be estimated from the number of noise events, without first determining the value of 1 PESD. Single and multiple-PESD signals can also be obtained precisely without any ambiguity. 

In this paper, we describe in Section \ref{sec2} the principle of this technique and demonstrate its use. In Section \ref{sec3}, we discuss and summarize the results.

\section{Principle of the Method and Measurements}
\label{sec2}
\subsection{Principle of the measurement for a single photoelectron}
\label{sec:principle}

When photoelectrons are ejected from the cathode of a PMT, an electrical potential of a few hundred volts is applied to attract them to the first dynode. The number of these primary photoelectrons follows a Poisson distribution $[3]$. %\cite{Bellamy}.
A photoelectron reaching the first dynode has a kinetic energy that depends on the electrical potential; a voltage of 100 V produces about 5 $\sim$ 10 secondary electrons from the dynode. Some electrons are backscattered from the dynode. The probability of scattering depends upon the dynode material; for copper, the probability is about 27 \% $[4, 5]$.%\cite{Sternglass, Razmik1}. 
The number of secondary electrons attracted from the first to the second dynode strongly affects the shape of the 1 PESD. 

\begin{figure}
  \centering
  \includegraphics[width=12cm]{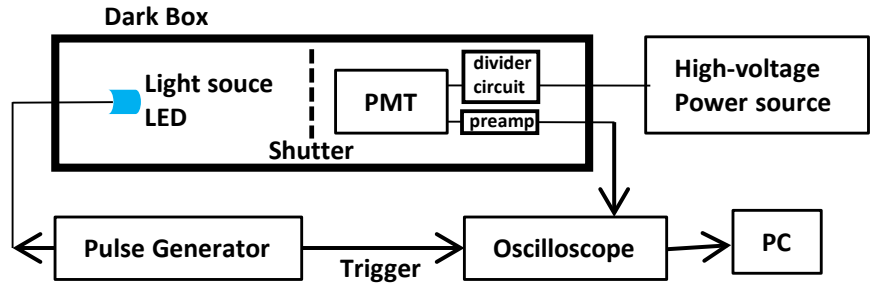}
  \caption{Schematic view of the experimental setup used to measure 1 PESD. A shutter is placed between the light source and the PMT.}
\label{fig:fig1}
\end{figure}
 
A schematic view of the experimental setup we used to measure 1 PESD is shown in Figure \ref{fig:fig1}. A light-emitting diode (LED) is used as the light source for this measurement. Electrical pulses are sent from a pulse generator to the LED, causing it to emit a weak flash lasting a few nanoseconds. The photons from the LED irradiate the PMT. For our measurements, we used a 1.5-inch PMT (HAMAMATSU R11920-100), which had been developed for the LST in the CTA $[6]$.%\cite{Toyama}. 
The signal from the PMT is amplified by a low-noise preamplifier with a gain of $\times$24 and is then transferred through a coaxial cable to an instrument such as an oscilloscope $[7]$.%\cite{PmtCta}.  

\begin{figure}
  \centering
  \includegraphics[width=8cm]{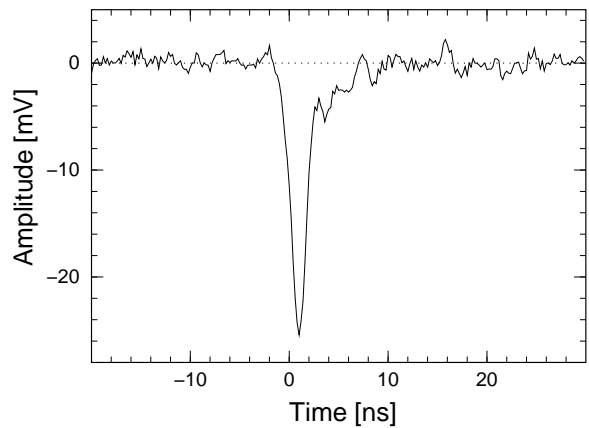}
  \caption{An example of a signal from the PMT. The signal was amplified by a factor of 24 by the preamplifier and was measured by an oscilloscope with a bandwidth of 300 MHz.
}
\label{fig:fig2}
\end{figure}

Trigger signals are also sent to the instrument from the pulse generator in order to synchronize the measurement with the LED flash. A typical signal is shown in Figure \ref{fig:fig2}, where the 3 ns pulse width can be clearly seen. 
The noise level in the signal is a few mV. By integrating the pulse, the signal can be obtained from the charge collected by the PMT anode. The integration time must be wide enough to contain all the signal from the PMT; in the present analysis, we used a 20 ns window. Since longer integration times increase the noise, the duration of the LED flash must be short. Moreover, one must select a PMT with a short pulse-width characteristic.

\begin{figure}[hbtp]
%  \centering
  \includegraphics[width=12cm]{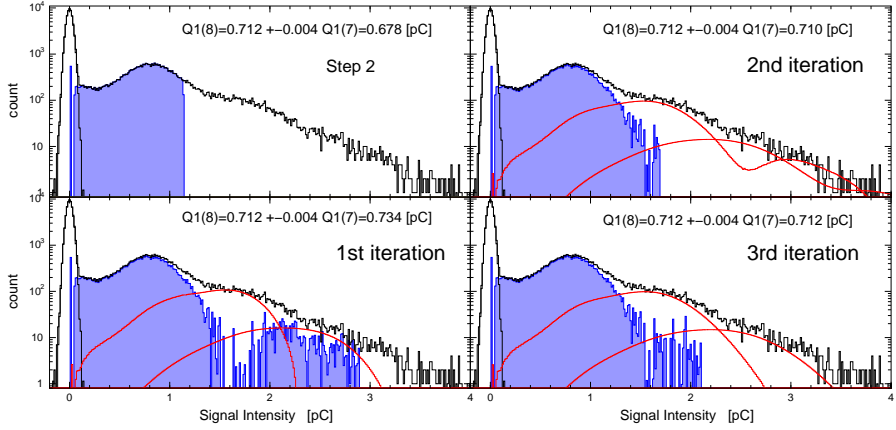}
  \caption{Distribution of signal from the PMT. A high voltage (1400 V) was applied to the PMT and the signal was amplified by a preamaplifier with a gain of 24. We measured 100,000 events with the shutter open. The signal from these events is shown by the black-line histogram. We also determined the quantity 0 PESD, or $n_0(i)$, by measuring 100,000 events with the shutter closed; this is also shown by a black-line histogram. The quantity 1 PESD, or $n_1(i)$, is shown by the shadowed histogram, and the distributions corresponding to 2 and 3 PESDs are shown by (red) solid lines. Our initial estimate of 1 PESD, obtained in Step 2 (see text), is shown in the upper left panel. The result of the first iteration in Step 4 is shown in the lower left panel, and the second and third iterations are shown in the right panels. Four iterations proved sufficient to converge the PESDs. The average signal corresponding to a single photoelectron, as obtained from Equations \ref{eq:07} and \ref{eq:08}, are also indicated in each panel.  
}
\label{fig:fig3}
\end{figure}
 
The distribution of the integrated signal is shown in Figure \ref{fig:fig3}. In this figure, the highest peak occurs around the signal level $q=0$. No photoelectrons reach the first dynode in these events, which just contain electrical noise. The bump around $q=0.8\; pC$ consists mainly of 1 PESD; i.e., the signal  corresponding to a single photoelectron is approximately $0.8\; pC$. Since the gain of the preamplifier is 24, the amplification factor of the PMT can be estimated approximately as $0.8\times 10^{-12}\; C/ (24\times 1.6 \times 10^{-19}\; C) \approx 2 \times 10^5$. An accurate measurement of this value and its associated statistical error is the main purpose of the present study. Events above 1 $pC$ contain two or more photoelectrons. Consequently 1 PESD can be found as the residual after 0 and two or more PESDs are subtracted from all events.

\subsection{Poisson distribution of the number of photoelectrons and 0 PESD}

In this section, we describe a procedure for estimating 1 PESD from the data. The goal of this analysis is to estimate 1 PESD and thus determine the amplification factor of the PMT, along with its statistical error.

As indicated in Figure \ref{fig:fig1}, a shutter is located between the LED and the PMT. Two measurements were performed using this shutter: in one, the PMT was illuminated with the shutter open, while in the second the shutter was closed. In both measurements, the signal from the PMT was integrated in synchronization with the LED flash event-by-event. We consider the charge of the integrated signal, represented by the symbol $q$ in units of C, to be the intensity of the signal. By making a histogram of the measured values of $q$, we obtained the distribution of the signal. Events are counted in bins of width of $\Delta q$ for each value of $q$. The counted value in the $i$th bin from the origin at $q = 0$ in the positive direction is represented by $n_{all}(i)$, and the $i$th bin in the negative direction is $n_{all}(-i)$. The quantity $i$ is the nearest integer of $q/\Delta q$, and the center of each bin is represented by $q(i)=i\times\Delta q$.

%\begin{equation}
%i =\frac{q}{\Delta q}
%\label{eq:01}
%\end{equation}

%The bin with $i=0$ spans in the range from $-\Delta q / 2$ to $\Delta q /2$, and the center of each bin is represented by $q(i) = i \times \Delta q$.

The total numbers of events with the shutter open and closed, respectively, are represented by $N_{all}$ and $N_{all}^{off}$. The number of $k$-photoelectron events -- i.e., number of events caused by exactly $k$ photoelectrons -- is represented by $N_k$. The number of $k$-photoelectron events in the $i$th bin is $n_k(i)$ and $n^{off}(i)$ is that obtained with the shutter closed. With these definitions, we have

\begin{eqnarray*}
N^{off}_{all} = N^{off}_0 = \sum_{i=-\infty}^\infty n^{off}(i) ,   
\label{eq:01_1}
\end{eqnarray*}
\begin{eqnarray*}
N_{all} = \sum_{i=-\infty}^\infty n_{all}(i) = \sum_{k=0}^\infty N_k
        = \sum_{k=0}^\infty \sum_{i=-\infty}^\infty n_k(i)   .
\label{eq:01_2}
\end{eqnarray*}

The average value of $k$ for all events is defined as $\langle k \rangle$; thus $\langle k \rangle$ is proportional to the brightness of the LED and to the photo-detection efficiency. It can be calculated from the following equation:

\begin{equation}
\langle k \rangle = \frac{\sum_{k=0}^\infty k \cdot N_k}{N_{all}} .
\label{eq:02}
\end{equation}

Since $N_k$ follows a Poisson distribution, it can be calculated from $\langle k \rangle$ by the following equation:

\begin{equation}
N_k = N_{all} \frac{\langle k \rangle ^k}{k!} e^{- \langle k \rangle}   .
\label{eq:03}
\end{equation}

Now we can determine $N_0$. Suppose that all events with $q<0$ are just noise. Then, we can obtain the result for zero photoelectrons -- i.e., 0 PESD -- from the data obtained with the shutter closed. We assume that no event caused by one or more photoelectrons makes a negative signal; this yields an estimated upper limit to $N_0$. Since the number of events with negative signal is negligibly small compared with the statistical error of $N_0$, we incorporate such negative signals into the systematic uncertainties instead of including them in the 1 PESD. This issue will be discussed further in Section \ref{sec:sys}.
Define the quantity $\alpha$ to be
\begin{equation}
\alpha = \frac{\sum_{i=-\infty}^{-1} n_{all}(i)}{\sum_{i=-\infty}^{-1} n^{off}(i)}   .
\label{eq:04}
\end{equation}
We thus obtain $N_0 = \alpha N^{off}_{0}$;
this estimate of $N_0$ causes the largest systematic uncertainty in the present analysis. Similarly, $n_0(i)$ -- or equivalently, 0 PESD -- can be estimated as $n_0(i) = \alpha n^{off}(i)$.

% The quantity 0 PESD or $n_0(i)$ can also be estimated roughly without measuring data with the shutter closed. Assuming that 0 PESD follows a Gaussian distribution, we can fit the distribution of $n(i)$ in the region $q<0$ to a Gaussian. The result can be considered approximately as 0 PESD. However, in this paper we estimate 0 PESD by using the data obtained with the shutter closed to avoid possible systematic error caused by such approximation.

Now the quantity $\langle k \rangle$ can be obtained from Equation \ref{eq:03} by setting $k=0$:

\begin{equation}
\langle k \rangle = ln\frac{N_{all}}{N_0}     .
\label{eq:05}
\end{equation}

%Once $N_0$ is estimated, $\langle k \rangle$ can then be calculated uniquely using Equation \ref{eq:05}. 
Substituting this result into Equation \ref{eq:03}, we obtain
\begin{equation}
N_k = \frac{\langle k \rangle ^k}{k!} N_0   .
\label{eq:06}
\end{equation}
That is, $N_k$ can be calculated for all $k$ using $N_0$. 
With $k=1$, we obtain $N_1=\langle k \rangle N_0$. Therefore Equation \ref{eq:06} can be rewritten as 
\begin{equation}
N_k = \frac{N_1^k}{k! N_0^{k-1}}    .
\label{eq:06-2}
\end{equation}

The average value of the signal corresponding to one photoelectron, $\langle Q_1 \rangle$, is defined as follows:
\begin{equation}
\langle Q_1 \rangle = \frac{\sum_{i=-\infty}^{\infty}n_1(i)q(i)}{N_1}   .
\label{eq:07}
\end{equation}
The quantity $\langle Q_1 \rangle$ can also be calculated from the average signal from all data divided by the average number of photoelectrons $\left( i.e., \langle Q_1 \rangle = \langle Q \rangle / \langle k \rangle \right)$. Thus, Equation \ref{eq:07} can also be written in the following alternative form:
\begin{equation}
\langle Q_1 \rangle = \frac{\sum_{i=-\infty}^{\infty}n_{all}(i)q(i)}{\langle k \rangle N_{all}}     .
\label{eq:08}
\end{equation}
Based on Equation \ref{eq:08}, we can thus obtain $\langle Q_1 \rangle$ without first determining 1 PESD; we only need to determine $N_0$. Comparing the results from Equations \ref{eq:07} and \ref{eq:08} also provides a useful check of the analysis. The gain of the PMT can then be obtained from $\langle Q_1 \rangle$ divided by the elementary charge and the gain of the preamplifier.

\subsection{Estimation of 1 PESD}

Given the value of $N_0$, 1 PESD can be determined as follows:

\begin{description}
\item[Step 1] We first determine the distribution of signal corresponding to one or more photoelectrons:
 
\begin{eqnarray*}
n_{k>0}(i) = n_{all}(i)-n_0(i)   .
\end{eqnarray*}
 
In this step, the initial values of $n_1(i)$ are set equal to $n_{k>0}(i)$ .

\item[Step 2]   We next integrate $n_1(i)$ from $i=-\infty$ up to the bin at which the integrated value is equal to  $N_1$, as calculated from Equation \ref{eq:06}. The last bin number for this integration is defined as $j$; that is, 
\begin{equation}
N_1 = \sum^{j}_{i=-\infty} n_1(i)        .
\label{eq:09}
\end{equation}

The value of $n_1(j)$ is adjusted so that Equation \ref{eq:09} is satisfied.

The bins with $i > j$ that contain one photoelectron are set to 0 (see the upper left panel in Figure \ref{fig:fig3}): 
\begin{eqnarray*}
n_1(i:i<j) = n_{k>0}(i) , \;\;\;\; n_1(i:i>j) = 0    .
\end{eqnarray*}
     
\item[Step 3] The distribution of two or more photoelectrons can be estimated from the superposition of $n_1(i)$. A signal with intensity $q(i)$ that is caused by two photoelectrons is represented by the convolution of two signals caused by single photoelectrons, with intensities $q(i^\prime)$ and $q(i-i^\prime)$. Therefore the probability distribution of the signal caused by two photoelectrons, $n_2(i)/N_2$, can be represented as follows: 
\begin{equation}
  \frac{n_2(i)}{N_2} = \sum^{\infty}_{i^\prime=-\infty} \frac{n_1(i^\prime)}{N_1}\frac{n_1(i-i^\prime)}{N_1}    .
\end{equation}
Using Equation \ref{eq:06-2}, we obtain $n_2(i)$ as
                
\begin{equation}
n_2(i) = \frac{1}{2!N_0}\sum^{\infty}_{i^\prime=-\infty} n_1(i^\prime)n_1(i-i^\prime)    .
\label{eq:10}
\end{equation}
In a similar way, we obtain the following equations:

\begin{eqnarray*}
n_3(i) = \frac{1}{3!(N_0)^2}\sum^{\infty}_{i^{\prime\prime}=-\infty} \sum^{\infty}_{i^\prime=-\infty} n_1(i^{\prime\prime})n_1(i^\prime)n_1(i-i^\prime-i^{\prime\prime})     ,
\end{eqnarray*}
 
\begin{eqnarray*}
n_4(i) = \frac{1}{4!(N_0)^3}\sum^{\infty}_{i^{\prime\prime\prime}=-\infty}\sum^{\infty}_{i^{\prime\prime}=-\infty} \sum^{\infty}_{i^\prime=-\infty} n_1(i^{\prime\prime\prime})n_1(i^{\prime\prime})n_1(i^\prime)n_1(i-i^\prime-i^{\prime\prime})     ,
\end{eqnarray*}
 
\begin{eqnarray*}
\cdots \cdots
\end{eqnarray*}

As long as Equation \ref{eq:09} is satisfied in Step 2, these equations ensure that the total number of events caused by $k$ photoelectrons is always correct; i.e., $N_k=\sum^{\infty}_{i=-\infty}n_k(i)$. Of course, in a real analysis, these calculations are performed over a limited range of $i$ and $k$. When $k$ is greater than 3, the calculation takes a long time, so it is impractical to conduct calculations for large $k$.
  
\item[Step 4]  The quantity $n_1(i)$ is now re-determined from  $n_k(i)$ as follows (see the lower left panel in Figure \ref{fig:fig3}):
\begin{eqnarray*}
n_1(i) = n_{k>0}(i) - \sum^{\infty}_{k=2} n_k(i)    .
\end{eqnarray*} 

\item[Step 5] Iterate from Step 2 through Step 4 until the distributions have all converged (the right panels in Figure \ref{fig:fig3}). In our analysis, we iterated these procedures four times since $\langle Q_1 \rangle$ converges to a constant value after four iterations.
\end{description}

\begin{figure}
  \centering
  \includegraphics[width=8cm]{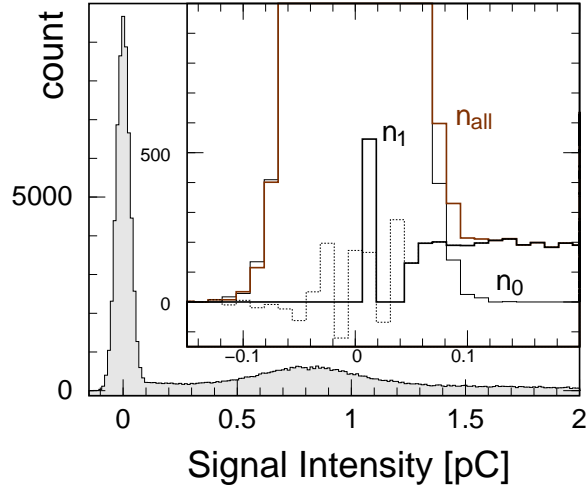}
  \caption{The distribution of signal. The distribution around $q=0$ is expanded in the inset panel. Both $N_{all}$ and $N^{off}_{all}$ are 100,000. The quantities $n_{all}(i)$, $n_0(i)$, and $n_1(i)$ are indicated by solid lines. Near the origin, we collapse the $n_1(i)$ into a single bin, as explained in Section \ref{sec:negativeBin}; the distribution prior to this change is indicated by the dashed line.}
\label{fig:fig4}
\end{figure}

\subsection{Handling of negative bins}
\label{sec:negativeBin}
In the analysis of real data, the statistical error from 0 PESD is relatively large. This causes negative values of $n_1(i)$ to occur around $q = 0$. The analysis cannot deal with negative bins, and we eliminated them as follows (see Figure \ref{fig:fig4}). 
Defining $\sigma$ as the RMS width of 0 PESD, we lump all the $n_1(i)$ with $q < +1.5 \sigma$ together in a single bin, with the bin number denoted by $i_0$. In other words, the width of the first bin of 1 PESD is enlarged to handle its large statistical error.
If $n_1(i_0)$ has a negative value, we artificially set it to 0. If the negative value is very large, this treatment produces large systematic errors. In such a case, however, there is likely to be some problem in the measurements. The charge $q(i_0)$ at the center of this bin is approximately defined as
\begin{eqnarray*}
q(i_0) = \frac{\sum^{1.5\sigma}_{i=-\infty} n_{k>0}(i)q(i)}{\sum^{1.5\sigma}_{i=-\infty} n_{k>0}(i)}     .
\end{eqnarray*} 
The value of $q(i_0)$ does not significantly affect the estimation of the PESDs, although $n_1(i_0)$ does make a significant contribution to the analysis, as discussed in Section \ref{sec:sys}. Except for $n_1(i_0)$, the values of $n_1(i)$ with $q(i)<+1.5 \sigma$ are  set to 0. 

\subsection{Statistical error}

It is not trivial to determine the statistical error of this analysis. In this section, we first estimate the statistical error of   $\langle Q_1 \rangle$ analytically from Equation \ref{eq:08}, which we denote as $\langle Q_1 \rangle ^{(8)}$. Using Equations \ref{eq:04} and \ref{eq:05}, $\langle k \rangle$ can be written in the form
\begin{eqnarray*}
\langle k \rangle &=& \ln N_{all} - \ln \alpha N^{off} \\
     &=& \ln N_{all} - \ln N^{off} - ln \sum_{i=-\infty}^{-1} n_{all}(i) + \ln \sum_{i=-\infty}^{-1} n^{off}(i)    .
\end{eqnarray*} 
With the approximation
\begin{eqnarray*}
\sum_{i=-\infty}^{-1} n^{off} \approx \frac{N^{off}}{2}   
\end{eqnarray*} 
the quantity $\langle k \rangle$ becomes 
\begin{eqnarray*}
\langle k \rangle \approx \ln \frac{N_{all}}{2} - \ln \sum_{i=-\infty}^{-1} n_{all}   .
\end{eqnarray*} 
Substituting this into Equation \ref{eq:08}, we obtain
\begin{eqnarray*}
\langle Q_1 \rangle^{(8)} = \frac{\sum_{i=-\infty}^{\infty}n_{all}(i)q(i)}
{\left( \ln \frac{N_{all}}{2} - \ln \sum^{-1}_{i=-\infty} n_{all}(i) \right) N_{all}}    .
\end{eqnarray*} 
In this equation, the only variable parameter is $n_{all}(i)$, and its error can be estimated as $\delta n_{all}=\sqrt{n_{all}(i)}$. This yields the following estimate of the error in $\langle Q_1 \rangle^{(8)}$:
\begin{small}
\begin{eqnarray*}
\left( \delta \langle Q_1 \rangle^{(8)} \right) ^2 
&=& \left( \delta n(i) \right)^2 \sum^{\infty}_{i=-\infty}\left( \frac{\partial \langle Q_1 \rangle^{(8)}}{\partial n_{all(i)}} \right) ^2 \\
&=& \frac{1}{N_{all}^2} \left( \sum^{-1}_{i=-\infty} \left( \frac{\partial}{\partial n_{all}(i)} \frac{\sum^{\infty}{i^\prime = -\infty} n_{all}(i^\prime)q(i^\prime)}{\ln \frac{N_{all}}{2} - \ln \sum^{-1}_{i^\prime = -\infty} n_{all}(i^\prime)} \right) ^2 n(i) + \frac{\sum^\infty_{i=0} q(i)^2 n_{all}(i)}{\left( \ln \frac{N_{all}}{2} - \ln \sum^{-1}_{i=-\infty} n_{all}(i) \right) ^2} \right) \\
&=& \frac{1}{\left( N_{all} \langle k \rangle \right)^2} 
\left( \sum^{-1}_{i=-\infty} \left( q(i) \langle k \rangle + \frac{\sum^\infty_{i^\prime = - \infty} n_{all}(i^\prime)q(i^\prime)}{\sum^\infty_{i^\prime = - \infty} n_{all}(i^\prime)} \right)^2 + \sum^\infty_{i=0} q(i)^2 n_{all}(i) \right)   .
\end{eqnarray*} 
\end{small}
The statistical error in $\langle Q_1 \rangle^{(7)}$ can be determined from a similar calculation, although it is more complicated. Instead, we estimate the statistical error of $\langle Q_1 \rangle^{(7)}$ by dividing the data into 10 fractions, determining the values of $\langle Q_1 \rangle^{(7)}$ for each fraction, and then estimating the error from the dispersion of these quantities. In this way, the values of $\langle Q_1 \rangle$ obtained from Equations \ref{eq:07} and \ref{eq:08} are found to be 
\begin{eqnarray*}
 \langle Q_1 \rangle^{(7)} &=& 0.712 \pm 0.008 \:\:\: [pC] , \\
 \langle Q_1 \rangle^{(8)} &=& 0.712 \pm 0.004 \:\:\: [pC] .
\end{eqnarray*}  
The error $\delta \langle Q_1 \rangle^{(7)}$ tends to be somewhat larger than $\delta \langle Q_1 \rangle^{(8)}$. This can be explained by the fact that the uncertainty in the shape of the distribution of $n^{off}(i)$ is included in $\delta \langle Q_1 \rangle^{(7)}$. The time variation of intensity of the LED may also affect this error. Note that $\langle Q_1 \rangle^{(8)}$ is consistent with $\langle Q_1 \rangle^{(7)}$. 

\begin{figure}
  \centering
  \includegraphics[width=12cm]{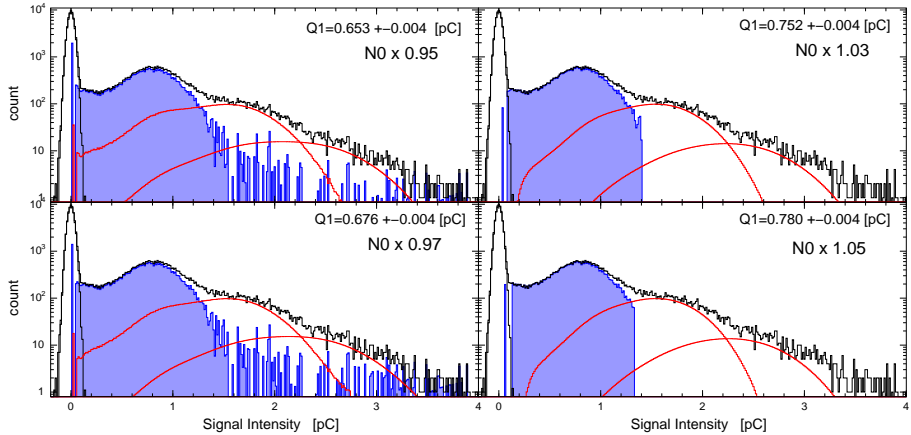}
  \caption{Effect of a systematic shift in $N_0$ on the estimate of 1 PESD using the same data as in Figure \ref{fig:fig3}. The value of $N_0$ used in Figure \ref{fig:fig3} is multiplied by 0.95, 0.97, 1.03, and 1.05 in proceeding, respectively, from the upper left to the lower right panels. }
\label{fig:fig5}
\end{figure}

\subsection{Systematic uncertainty}
\label{sec:sys}
The largest systematic uncertainty in this analysis is caused by the number of low-intensity events, because the distribution around $q=0$ is sensitive to environmental effects on the measurements. For example, a small current to the shutter may cause additional pickup noise. 

The effect of an incorrect estimate of $N_0$ can be seen in Figure \ref{fig:fig5}. Since $\langle k \rangle$ depends on $N_0$, the value of $\langle Q_1 \rangle$ is affected by the estimate of $N_0$. As shown in Figure \ref{fig:fig5}, a 3 \% change in $N_0$ causes a 5 \% shift in  the value of $\langle Q_1 \rangle$ and distorts the estimated 1 PESD from the Poisson-like distribution shown in Figure \ref{fig:fig3}. Therefore, if $N_0$ differs by 3 \% or less, the systematic uncertainty in the value of $\langle Q_1 \rangle$ is estimated as 5 \%. 

Another source of systematic uncertainty is the value of $n_1(i_0)$,  into which the data around $q=0$ is collapsed (Section \ref{sec:negativeBin}). While the value of $q(i_0)$ does not significantly affect the analysis, the number of events in this bin does produce a systematic uncertainty, which affects the estimate of $\langle Q_1 \rangle^{(7)}$. If this value were different by as much as 30 \%, the estimate of $\langle Q_1 \rangle^{(7)}$ would only change by 3 \%. 

Adding these values quadratically, the systematic uncertainties in the value of $\langle Q_1 \rangle$ can thus be estimated as 6 \%. Of course, this uncertainty depends on the setup of the measurements, and it may be possible to reduce it.

Additional sources of systematic uncertainty include the dark current and the after-pulse in the PMT. The dark current is mainly caused by electrons liberated by thermal fluctuations in the cathode, which are emitted randomly at a rate less than approximately 1 MHz. The after pulse is caused by secondary electrons, which are multiplied in the PMT and occasionally collide with gas atoms or molecules, creating positive ions in the PMT. The positive ions are attracted to the cathode by the electric field, ejecting additional electrons that produce a relatively large signal. The rate of the after-pulse depends on the quality of the PMT; in general, there is less than a 1 \% chance to produce an after-pulse by one photoelectron.
These false signals can be neglected as long as the integration time is of the order of 20 ns.

\begin{figure}
  \centering
  \includegraphics[width=12cm]{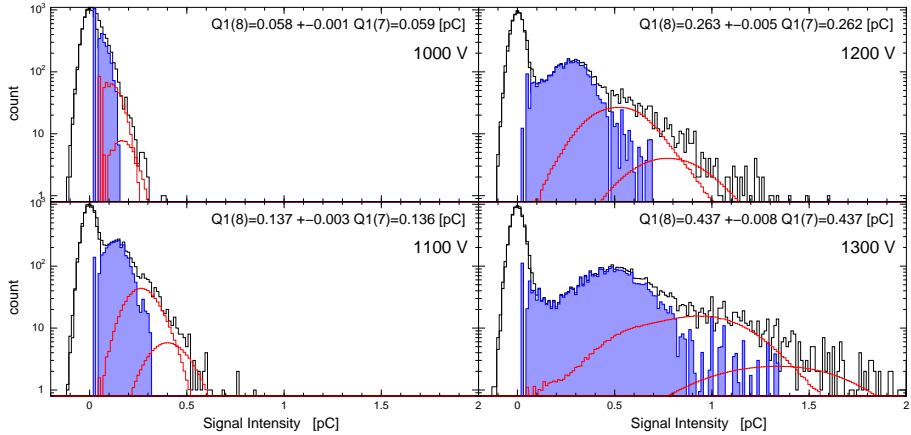}
  \caption{Estimates of 1 PESD obtained with various PMT gains. The average signal from single photoelectrons $\langle Q_1 \rangle$, as estimated from Equations \ref{eq:07} and \ref{eq:08}, are also indicated with applied high voltage in each panel. 10,000 events for both of shutter opened and closed are used in these measuements.
}
\label{fig:fig6}
\end{figure}

\section{Discussion and Summary}
\label{sec3}
\subsection{Discussion}
The number of secondary electrons created at the first dynode is distributed according to a Poisson distribution. This distribution makes the largest contribution to the shape of 1 PESD. However, as can be seen in the lower right panel of Figure \ref{fig:fig3}, the 1 PESD departs from a Poisson distribution at low intensity. Single-photoelectron events are increased in the low-intensity region, as compared with a simple Poisson distribution. This can be explained by backscattering of the photoelectrons at the first dynode as mentioned in Section \ref{sec:principle}. In such a case, the photoelectron retains some of its energy, reducing the number of secondary electrons that proceed to the second dynode. Consequently, the signals from such events decrease. One of the advantages of the present analysis is that these small signals are precisely included in the determination of $\langle Q_1 \rangle$ and 1 PESD. 

When the gain of the PMT is small, or if the noise level is high compared with the single-photoelectron signal, it becomes difficult to discriminate between the 1 PESD and 0 PESD signals. To test for this condition, we estimated 1 PESD using several values of the high voltage on the PMT. The resulting 1 PESIDs, corresponding to PMT gains from $10^4$ to $10^5$, are shown in Figure \ref{fig:fig6}. The statistical error of the estimated $\langle Q_1 \rangle$ decreases with increasing PMT gain, as shown in this figure, although the errors are about 2 \% for all high voltages. When the high voltage is 1000 V, the 1 PESD cannot be discriminated by eye from the noise component (0 PESD). However, even in this case, both 1 PESD and the gain of the PMT can be clearly obtained using this analysis so long as $N_0$ is determined correctly.

\subsection{Summary}
We have developed a procedure to estimate the average signal produced by a single photoelectron, along with its distribution, which is one of the most fundamental parameters required for measuring the intensity of weak light using a photon detector such as a PMT. In this method, we utilize the fact that the number of photoelectrons reaching the first dynode follows a Poisson distribution. Comparing the distributions of signals from the PMT obtained with the light source turned on and off (or, equivalently, with the shutter open and closed), we can estimate the number of events with zero photoelectrons, denoted by $N_0$. Once $N_0$ and the total number of events, denoted by $N_{all}$, are determined, the average number $\langle k \rangle$ of photoelectrons reaching the first dynode can be calculated uniquely from the Poisson distribution. Then the number of events $N_k$ for each number of photoelectrons can also be determined. With this method, we can obtain the gain of the PMT and its associated error without first determining the distribution of signal for events caused by single photoelectrons, which we define as 1 PESD. The 1 PESD distribution can also be estimated from the fact that a $k$-PESD distribution is the superposition of $k$ times the 1 PESD distribution. 

Based on this method, we determined the average signal due to a single photoelectron, $\langle Q_1 \rangle$, to within a 1 \% statistical error when the gain of PMT was $10^5$ and the number of measured events was 100,000.  This statistical error is inversely proportional to the square of the number of events. Even though the 0 and 1 PESD distributions cannot be discriminated by eye if the gain of the PMT is low and/or the noise level is high, the quantity $\langle Q_1 \rangle$ can be obtrained, along with its statistical error, using this method. This procedure does not require fitting routines nor any assumptions about the shape of the 1 PESD distribution. The shape of 1 PESD can also be determined experimentally using this method. Systematic uncertainties in this analysis are mainly caused by the determination of $N_0$, which we  estimated as 6 \%. The most important factor in this measurement is thus a precise measurement of 0 PESD.

The 1 PESD distribution we obtained shows that events at small signal levels exceed the numbers expected from a Poisson distribution. As the quality of PMTs improve, the detailed structure of 1 PESD will become clearer in the measurements. Consequently, precise calibration is required to make the best use of the quality of the detector. This requirement led to the present study. 

In summary, we have developed a procedure to estimate the average signal produced by a single photoelectron, $\langle Q_1 \rangle$, and to obtain the gain of the PMT. The most important parameter turns out to be the number of events caused by zero photoelectrons, namely noise events. Direct measurement of the signal produced by a single photoelectron proves to be unnecessary.

\section*{Acknowledgments}
The authors thank the CTA collaboration for their assistance. 
%This work is supported by the Ministry of Education, Culture, Sports, Science and Technology, 
%Japan, a Grant-in-Aid for Specially Promoted Research, 24000004-01, 2011, and
%by a Grant-in-Aid for Challenging Exploratory Research, 21654036, 2015. 
This work is supported by JSPS KAKENHI Grant Number 24000004-01, 15K13489, and 17H06131.
Furthermore, it was partly supported by the joint research program of the Institute of Cosmic Ray Research (ICRR), University of Tokyo. The authors thank Enago (www.enago.jp) for the English language review.
\vspace{0.5cm}
\\
This paper has gone through internal review by the CTA Consortium.\\ \\

%\section{Bibliography styles}
%\section*{References}
%\bibliography{mybibfile}
{\bf References}
\\ \\
%¥begin{thebibliography}{9}
%\bibitem{Hamamatsu} Hamamatsu Photonics,''Photomultiplier Tubes Basics and Applications,'' 2007.
%\bibitem{CTA} a
%\bibitem{Bellamy} b
%\bibitem{Sternglass} c
%\bibitem{Razmik1} d
%\bibitem{Toyama} d
% \bibitem{PmtCta} f
%¥end{thebibliography}
$[1]$ Hamamatsu Photonics K. Photomultiplier Tube Basics and Applications, \\
      \hspace{5mm}2007 \\
$[2]$ CTA consortium, Design concept for the Cherenkov telescope array,\\ 
      \hspace{5mm}Exp. Astron. 32 (2011) 192-316. arxiv:1008.3703\\
$[3]$ E.B. Bellamy, B. Bellettini, J. Budagov, F. Cervelli, I. Chirikov-Zorin,\\
      \hspace{5mm}M. Incagli, D. Lucchesi, C. Pagliarone, S. Tokar, F Zetti, Absolute\\
      \hspace{5mm}calibration and monitoring of a spectrometric channel using a\\
      \hspace{5mm}photomultiplier, Nucl. Instrum. Methods Phys. Res. A 339 (1994) 468-476.\\
$[4]$ E.J. Sternglass, Backscattring of kilovolt electrons from solids, Phys. Rev. \\
      \hspace{5mm}95 (1954) 245-358.\\
$[5]$ R. Mirzoyan, E. Lorenz, On the calibration accuracy of light sensors in \\
     \hspace{5mm}atmospheric cherenkov, fluorescence and neutrino, in: 25th ICRR in Durban, \\
     \hspace{5mm}vol. 7, 1997, p. 265\\
$[6]$ T. Toyama, R. Mirzoyan, et al., for the CTA consortium, Novel photo \\
     \hspace{5mm}multiplier tubes for the Cherenkov telescope array project, in: 33th ICRC \\
     \hspace{5mm}in Rio de Janeiro, 2013, aXiv1307.5463\\
$[7]$ R. Mirzoyan, D. Mueller, Y. Hanabata, J. Hose, D. Menzel, M. Takahashi, \\
     \hspace{5mm}M. Teshima, T. Toyama, T. Yamamoto, Evaluation of photo multiplier tube \\
     \hspace{5mm}for the Cherenkov telescope array, Nucl. Instrum. Methods Phys. Res. A \\
     \hspace{5mm}824 (2016) 640-641.\\

\end{document}